\documentclass[prl,reprint,
preprintnumbers]{revtex4-2}
\pdfoutput=1
\usepackage{hyperref}
\hypersetup{hypertexnames=false}
\usepackage[T1]{fontenc}

\usepackage{mathrsfs}

\usepackage[inline]{enumitem}

\usepackage{amsmath,amssymb,amsfonts,amsxtra,mathrsfs,graphics,graphicx,amsthm,epsfig,bm,longtable,float,color,tikz,mathtools,xfrac,footnote}
\usepackage{dsfont}
\usepackage{lipsum}
\usepackage{textgreek}
\usetikzlibrary{decorations.pathmorphing}
\usetikzlibrary{decorations.markings}
\usetikzlibrary{quotes,arrows.meta}
\usetikzlibrary{arrows,decorations.markings,calc,fadings,decorations.pathreplacing,patterns,decorations.pathmorphing,positioning}
\usepackage{tikz-cd}

\newcommand{\ii}{\mathrm{i}}

\newcommand{\nn}{\nonumber}

\newcommand{\be}{\begin{equation}} \newcommand{\ee}{\end{equation}}
\newcommand{\bea}{\begin{equation} \begin{aligned}} \newcommand{\eea}{\end{aligned} \end{equation}}

\newcommand{\e}{\mathrm{e}}

\usepackage{relsize}

\newcommand{\wt}{\widetilde}
\newcommand{\wh}{\widehat}

\newcommand{\smallstrut}{\rule{0pt}{.6em}}

\DeclareMathOperator{\re}{Re}
\DeclareMathOperator{\im}{Im}

\DeclareMathOperator{\Li}{Li}

\newcommand{\cL}{\mathcal{L}}

\newcommand{\cN}{\mathcal{N}}

\newcommand{\cW}{\mathcal{W}}

\usepackage[bbgreekl]{mathbbol}
\DeclareMathSymbol\bbDelta  \mathord{bbold}{"01}

\usepackage{orcidlink}

\newcommand{\Nh}{\wh N}
\newcommand{\lamh}{\wh\lambda}
\newcommand{\CW}{C}
\newcommand{\Cint}{C_{\rm int}}
\newcommand{\sthree}{S^3}
\usetikzlibrary{shapes.geometric}
\graphicspath{{figs/}}

\newcommand{\Wt}{\wt{\cW}}
\newcommand{\Cltwo}{\mathrm{Cl}_2}
\newcommand{\rd}{\mathrm{d}}
\newcommand{\Hgen}{\mathcal{H}}

\begin{document}

\title{Exact Modular Completion of the ABJM Effective Twisted Superpotential}

\author{Seyed Morteza Hosseini\,\orcidlink{0000-0001-8205-400X}}
\affiliation{Centre for Theoretical Physics, Department of Physics and Astronomy, Queen Mary University of London, London E1 4NS, UK}

\begin{abstract}
The effective twisted superpotential governs the supersymmetric partition functions of three-dimensional $\mathcal{N}=2$ theories in the Cardy limit and, at large $N$, the gravitational blocks that are glued into the entropy function of supersymmetric $\mathrm{AdS}_4$ black holes. We determine its on-shell structure for $\mathrm{U}(N)_k\times\mathrm{U}(N)_{-k}$ ABJM theory at the universal twist, using a machine-learning discovery pipeline---physics-informed symbolic regression with integer-relation detection---applied to Bethe-vacuum data of up to $800$ digits. Its perturbative part terminates after two terms, a shifted-rank $3/2$-power and an $N$-independent constant map, which we obtain in closed form for every integer level. Its type-IIA genus expansion admits a closed coefficient formula at arbitrary genus, involving binomial and Bernoulli-number terms, and is asymptotic. For $k=1,2,4$ the entire finite-$N$ remainder is a single absolutely convergent divisor-sum $q$-series, obtained by applying a first-order differential operator to the Eichler integral of a weight-four Eisenstein series on $\Gamma_0(2)$ or $\Gamma_0(4)$. This closed form reproduces the data to $\sim\!10^{-797}$.
\end{abstract}

\maketitle

\emph{Introduction.}---%
Supersymmetric partition functions of three-dimensional $\cN = 2$ theories factorize into holomorphic blocks labelled by massive vacua~\cite{Pasquetti:2011fj,Beem:2012mb}. In the Cardy limit, $\omega\to0$, each block is governed by the effective twisted superpotential $\Wt$ evaluated at a Bethe vacuum, $B_\alpha\sim\exp[-\tfrac1{2\omega}\Wt(u_\alpha,\Delta)]$, where $\omega$ is a background parameter and $\Delta$ denotes the flavor chemical potentials. One object therefore controls the Bethe-ansatz representation of the topologically twisted index~\cite{Nekrasov:2009uh,Benini:2015noa}, the Cardy limits of superconformal and refined indices~\cite{Choi:2019dfu,Hosseini:2022vho}, and supersymmetric partition functions on any closed oriented Seifert manifold~\cite{Closset:2018ghr}, the squashed sphere included. Holographically, the dominant block is identified at large $N$ with the gravitational blocks that are glued into the entropy function of rotating, supersymmetric $\mathrm{AdS}_4$ black holes carrying generic electric and magnetic charges~\cite{Hosseini:2019iad}. This is reason enough to study $\Wt$ of $\mathrm{U}(N)_k\times\mathrm{U}(N)_{-k}$ ABJM theory~\cite{Aharony:2008ug}---the worldvolume theory of $N$ M2-branes probing $\mathbb{C}^4/\mathbb{Z}_k$---as an observable in its own right.

The $S^3$ free energy $F_{\sthree}$ was obtained at planar level, with its $N^{3/2}$ scaling and worldsheet-instanton corrections, from the lens-space matrix model~\cite{Marino:2009jd,Drukker:2010nc}; its perturbative all-genus expansion was resummed into an Airy function in~\cite{Fuji:2011km}, and subsequently rederived and generalized through the Fermi-gas reformulation~\cite{Marino:2011eh}. The twisted index at the universal twist has a perturbatively exact form controlled by a shifted rank~\cite{Bobev:2022jte}. Their \emph{nonperturbative} completions are another matter: for $F_{\sthree}$ the exponentially small corrections assemble the full topological string on local $\mathbb{P}^1\times\mathbb{P}^1$, an asymptotic transseries of worldsheet, membrane and bound-state instantons~\cite{Drukker:2010nc,Hatsuda:2012dt,Calvo:2012du,Hatsuda:2013gj,Hatsuda:2013oxa,Hatsuda:2015gca}. Whether $\Wt$ inherits that complexity has not been settled; nor has its $N$-independent constant map---the one piece of the perturbative answer no scaling argument fixes---or its exponentially small tail.

Conventional tools are ill-suited to the second question. A matrix-model expansion yields the perturbative series but not resummed instanton coefficients; a numerical fit yields approximate coefficients but no certificate of exactness. What is needed is the reverse of interpolation: given a number known to hundreds of digits, decide whether it is a specific rational combination of a small set of transcendental constants. This is the province of machine-learning methods for exact formula discovery---physics-informed symbolic regression~\cite{Udrescu:2019mnk,Cranmer:2023pysr}, sparse high-precision regression, and integer-relation detection~\cite{ferguson1999analysis,Raayoni:2021ram}---and it is our instrument.

Our results are as follows. (i) The perturbative $\Wt$ closes after a shifted-rank $3/2$-power and a constant, with no intermediate powers of $1/N$; high-precision reconstruction identifies parity-dependent finite Clausen formulas and a common convergent integral. The Supplemental Material reduces the integral exactly to a rational-argument Euler sum and reports an independent high-precision comparison of the Clausen, Euler, and integral representations. The root-of-unity evaluation of the Euler sum, which proves the formulas for every positive integer level, is deferred to the companion paper~\cite{HosseiniTTI}. (ii) Its fixed-$\lambda$ type-IIA genus expansion admits a closed coefficient formula at arbitrary genus, involving binomial and Bernoulli-number terms, and is asymptotic. (iii) At $k=1,2,4$ the complete finite-$N$ remainder is an absolutely convergent divisor-sum $q$-series generated by a first-order differential operator acting on a weight-four Eichler integral of the Eisenstein series $E_4(Q)-L^2E_4(Q^L)$ on $\Gamma_0(L)$, with $L=2$ for $k=1,2$ and $L=4$ for $k=4$.

\emph{Setup and conventions.}---%
ABJM theory has gauge group $\mathrm{U}(N)_k\times\mathrm{U}(N)_{-k}$ and chiral multiplets $A_{1,2}$ in the bifundamental $(\mathbf{N},\overline{\mathbf{N}})$ and $B_{1,2}$ in the conjugate bifundamental $(\overline{\mathbf{N}},\mathbf{N})$, coupled through the superpotential $W=\varepsilon^{ab}\varepsilon^{cd}\,\mathrm{Tr}\,A_aB_cA_bB_d$; its $\cN=6$ supersymmetry is enhanced to $\cN=8$ at $k=1,2$ by monopole operators~\cite{Aharony:2008ug,Gustavsson:2009pm}. The index $a=1,\dots,4$ runs over the four chiral fields $(A_1,A_2,B_1,B_2)$ and the Abelian flavor symmetries $\mathrm{U}(1)^4\subset\mathrm{SO}(8)$, with chemical potentials $\Delta_a$;
the superpotential enforces $\sum_{a=1}^{4}\Delta_a=2\pi$. Let $u_i,\wt u_i$ ($i=1,\dots,N$) be the angular Coulomb-branch (holonomy) variables, defined modulo $2\pi$.
The effective twisted superpotential is~\cite{Benini:2015eyy}
\bea
\Wt&=\sum_{i=1}^{N}\Big[\frac{k}{2}(\wt u_i^{2}-u_i^{2})
-2\pi(\wt n_i\wt u_i-n_i u_i)\Big] \\
&+\sum_{i,j=1}^{N} \sum_{\sigma=\pm1}\sigma
\sum_{a\in I_\sigma}
\Li_2 \big(\e^{\ii(\wt u_j-u_i+\sigma\Delta_a)}\big) \, ,
\label{eq:Wdef}
\eea
with $I_{-}=\{1,2\}$, $I_{+}=\{3,4\}$, and $\Li_s$ the principal analytic continuation of the polylogarithm, normalized by $\Li_s(z)=\sum_{n\ge1}z^n/n^s$ for $|z|<1$; the principal dilogarithm has its cut on $[1,\infty)$. Since $\Wt$ is multivalued, its critical points exist only after a lift: the Bethe equations are $\partial_{u_i}\Wt=\partial_{\wt u_i}\Wt=0$, and the integers $n_i,\wt n_i\in\mathbb{Z}$ in \eqref{eq:Wdef} parametrize the resulting angular ambiguities. We adopt $n_i=\wt n_i=-i$, with the eigenvalues labelled by increasing imaginary part. This is the assignment for which the long-range forces between eigenvalues cancel and the twisted superpotential is known to scale as $N^{3/2}$ at large $N$~\cite{Benini:2015eyy}. Every statement below---in particular the value of the constant map $\CW(k)$---refers to this dominant-vacuum, principal-branch prescription.

We work at the universal twist $\Delta_a=\pi/2$ and write $\Wt(N,k)$ for the on-shell value at the dominant vacuum. On that vacuum the data return $\re\Wt(N,k)=0$ identically---to the full working precision, at every level and rank we computed---so only the imaginary part carries information. The Bethe equations are solved numerically in two independent ways: by reading them as equilibrium conditions for interacting particles in the complex plane and relaxing an initial configuration by gradient flow~\cite{Herzog:2010hf,Benini:2015eyy}, refined by Newton iteration; and by the continuation scheme of Ref.~\cite{Hosseini:2025jxb}. The two agree. We produce $200$-digit solutions for $1\le k\le50$, covering even ranks $N\le800$ at $k=1$ and $N\le500$ for $2\le k\le50$, together with odd ranks $N\le299$ at every level, and $800$-digit solutions for $k=1,2,4$ with $2\le N\le100$. All discovery below reads these decimal strings, never machine floating point.

The natural variables are the shifted rank, the instanton action and the nome,
\be
\Nh=N-\frac{k}{24}+\frac{2}{3k}\,,\qquad t=2\pi\sqrt{\frac{2\Nh}{k}}\,,\qquad q=\e^{-t} \, ,
\label{eq:shift}
\ee
where \eqref{eq:shift} is recovered from the data; all ranks used below have $\Nh>0$.
We write $\sigma_3(n)=\sum_{d\mid n}d^{3}$ (with $\sigma_3(x)=0$ for $x\notin\mathbb{Z}_{>0}$), $\Cltwo(\theta)=\im\Li_2(\e^{\ii\theta})=\sum_{n\ge1}\sin(n\theta)/n^2$ for the Clausen function, $G=\Cltwo(\pi/2)$ for Catalan's constant, and $\lambda=N/k$, $\lamh=\lambda-\tfrac1{24}$ for the 't~Hooft coupling.

\emph{The perturbative parent and its constant map.}---%
The perturbative part of $\im\Wt(N,k)$ terminates, at fixed $k$, after two terms:
\be
\im\Wt(N,k)\big|_{\rm pert}=\frac{\pi^{2}}{3}\sqrt{\frac{k}{2}}\;\Nh^{3/2}+\CW(k) \, .
\label{eq:parent}
\ee
We verify directly that the residual left by \eqref{eq:parent} is exponentially small in $\Nh^{1/2}$---of order $q$, not of order any power of $1/N$---so \eqref{eq:parent} is exact to all orders in the fixed-$k$ perturbative expansion. This already contrasts with the sphere, where $F_{\sthree}$ requires an infinite tower of $1/N$ corrections resumming into an Airy function~\cite{Fuji:2011km,Marino:2011eh}; the two observables share the $3/2$ scaling and a rank shift, but \eqref{eq:shift} is the universal-twist shift~\cite{Bobev:2022jte} and differs from the Airy one by $1/k$. At leading order, in the normalization of \eqref{eq:Wdef}, $\im\Wt|_{\rm pert}=\tfrac{\pi}{2}F_{\sthree}$~\cite{Hosseini:2016tor}; at finite $N$ they are distinct observables.

Once the shifted rank \eqref{eq:shift} and the leading $3/2$-power are fixed, the constant map is the only $N$-independent datum left. For each integer level we form $\CW^{\rm data}(N,k)=\im\Wt(N,k)-\tfrac{\pi^{2}}{3}\sqrt{k/2}\;\Nh^{3/2}$, which approaches its limit exponentially in $\Nh^{1/2}$ at the first-instanton rate. We take the value at the largest available rank as the estimator for $\CW(k)$, and the gap between the two largest ranks as a conservative bound on the leftover contamination. At $k=1$ the estimator moves by only $4.7\times10^{-108}$ between $N=798$ and $N=800$.

The plateau is a single number, so identifying it is an integer-relation problem, and the choice of basis is the only modelling step. We use $\zeta(3)/\pi$, $G$, and $\Cltwo(2\pi j/k)$ for $j=1,\dots,k-1$. Clausen values are expected because $\Wt$ is built from dilogarithms and $\im\Li_2(\e^{\ii\theta})=\Cltwo(\theta)$; their arguments are placed at roots of unity because the level enters the Bethe equations as an integer; $\zeta(3)/\pi$ is indicated by the data themselves, the plateau growing as $-k^{2}\zeta(3)/16\pi$; and Catalan's constant is listed separately because $G=\Cltwo(\pi/2)$ belongs to the root-of-unity set only when $4\mid k$. The choice is an ansatz, but a falsifiable one---PSLQ returns nothing below the height bound when a constant is missing, and relations that dissolve under pruning when the basis is redundant. The raw list is also degenerate---$\Cltwo(2\pi(k-j)/k)=-\Cltwo(2\pi j/k)$, $\Cltwo(\pi)=0$, and the duplication formula $\Cltwo(2\theta)=2\Cltwo(\theta)+2\Cltwo(\theta+\pi)$ relates the survivors---so PSLQ is first run on the basis values alone, pruning the list until no relation remains below the height bound, before the data are touched. A relation found on a level's plateau is then accepted only if its height is low and it survives two perturbations: cutting the working precision by one quarter, and deleting the basis elements it does not use.

The method's reach hinges on a competition: the digits that the plateau supplies versus the precision that a certificate demands. Supply declines monotonically with level as the instanton floor rises---$106$ digits at $k=1$, down to just $8$ at $k=50$. Certification of an $n$-term relation with coefficient height $H$ requires roughly $n\log_{10}H$ digits~\cite{ferguson1999analysis}; crucially, $n$ and $H$ are determined by the surviving Clausen span at each level $k$.
Successful certification continues unbroken through $k=8$, then only $k=10$ and $k=12$ survive---their Clausen spans narrowing to dimensions two and one---while $k=9,11$, and all $k\ge13$ fall short.

Reading the certified relations as a function of $k$ is the one step done by hand. Four odd levels and three even ones suffice: they split on the parity of $k$ and assemble into closed formulas. For odd $k=p$,
\bea
\CW(p)&=-\Big(\frac{p^{2}}{4}+\frac{3}{8p}\Big)\frac{\zeta(3)}{\pi}+(-1)^{\frac{p+1}{2}}G \\
&\;-\frac{3}{8}\sum_{j=1}^{p-1}\big(1-(-1)^{j}\big)\,j\,\Cltwo\!\Big(\frac{2\pi j}{p}\Big) \\
&\;-\frac{1}{4}\sum_{j=1}^{p-1}(-1)^{j}\,j\,\Cltwo\!\Big(\frac{4\pi j}{p}\Big) \, ,
\label{eq:clausen-odd}
\eea
while for even $k=2p$,
\bea
\CW(2p)&=-\Big(p^{2}-\frac{1+7(-1)^{p}}{4p}\Big)\frac{\zeta(3)}{\pi} \\
&\;-\sum_{j=1}^{p-1}\frac{3-(-1)^{p+j}}{2}\,j\,\Cltwo\!\Big(\frac{2\pi j}{p}\Big) \, .
\label{eq:clausen-even}
\eea
These are arithmetically reconstructed---exact rationals, not fits. Two checks close the loop. The certificates at $k=8$, $10$ and $12$, none of which entered the assembly, coincide integer by integer with what \eqref{eq:clausen-odd}--\eqref{eq:clausen-even} return in the same pruned basis; and the formulas predict the plateau at every one of the forty-three levels $k=8,\dots,50$ withheld from their construction, each to the instanton floor of that level. Table~\ref{tab:CW} of the Supplemental Material collects the resulting closed forms through $k=8$; the exceptional levels are elementary, $\CW(1)=-G-\tfrac{5\zeta(3)}{8\pi}$, $\CW(2)=-\tfrac{5\zeta(3)}{2\pi}$, and $\CW(4)=-\tfrac{3\zeta(3)}{\pi}$.

The same parity-dependent expressions admit a common integral representation.
A companion paper~\cite{HosseiniTTI} proves that, for every positive integer $k$,
\be
\CW(k)=-\frac{k^{2}\zeta(3)}{16\pi}+\frac{2k}{\pi}\!\int_{0}^{\infty}\!\!\log(2\cosh x)\,\log\!\big(1-\e^{-kx}\big)\,\rd x \, .
\label{eq:CW}
\ee
The Supplemental Material establishes absolute convergence and reduces the
right-hand side of \eqref{eq:CW} exactly to an absolutely convergent
alternating Euler sum in the generalized harmonic numbers $\Hgen_{2m/k}$,
where $\Hgen_z=\psi(1+z)+\gamma$ for $z\ge0$, with $\psi$ the digamma function
and $\gamma$ the Euler--Mascheroni constant. The remaining root-of-unity
evaluation of that rational-argument Euler sum is deferred to
Ref.~\cite{HosseiniTTI}. We supplement the analytic reduction with a three-way
numerical test: at $70$-digit working precision, the finite Clausen
expressions, the Euler representation, and direct quadrature of \eqref{eq:CW}
agree for every $1\le k\le50$---independently of the Bethe-vacuum
data---with largest pairwise discrepancy below $7\times10^{-69}$. At $k=2$ the
Euler series reduces to the classical alternating harmonic sum and proves
\eqref{eq:CW} directly at that level. Equation~\eqref{eq:parent} with
\eqref{eq:clausen-odd}--\eqref{eq:clausen-even} matches the data plateau down
to the instanton floor: the residual at the largest available rank is
$1.3\times10^{-107}$ at $k=1$ ($N_{\max}=800$), $9.2\times10^{-37}$ at $k=5$,
and $7.4\times10^{-9}$ at $k=50$ (both $N_{\max}=500$); the agreement degrades
with $k$ as $q$ grows.

\emph{All-genus coefficient formula.}---%
At fixed $\lambda$ the parent \eqref{eq:parent} organizes into a string-genus expansion
\be
\im\Wt(N,k)\big|_{\rm pert}=-\sum_{h\ge0}(2\pi\ii\lambda)^{2h-2}\,\cW_h(\lambda)\,N^{2-2h} \, .
\label{eq:Wgenus}
\ee
Expanding \eqref{eq:parent} at large $k$ with $\lambda$ fixed,
\be
\cW_0=\frac{2\sqrt2\pi^{4}}{3}\lamh^{3/2}-\frac{\pi\zeta(3)}{4} \, , \quad
\cW_1=-\frac{\pi^{2}}{3\sqrt2}\lamh^{1/2}+\frac{\pi}{3}\log2 \, ,
\label{eq:W01}
\ee
and, for every $h=n+1\ge2$,
\bea
\cW_{n+1}&=\frac{(-1)^{n+1}\sqrt2}{2^{n}3^{n+2}\pi^{2n-2}}\binom{3/2}{n+1}\lamh^{\frac12-n}+\omega_n \, , \\
\omega_n&=-\frac{2^{2n+1}(2^{2n}-1)\,\pi\,|B_{2n}B_{2n+2}|}{n\,(2n+2)!} \, ,
\label{eq:Wgen}
\eea
with $B_{2n}$ the Bernoulli numbers,\footnote{We normalize them by $x/(\e^{x}-1)=\sum_{m\ge0}B_m\,x^{m}/m!$, so that $B_2=1/6$, $B_4=-1/30$ and $B_6=1/42$.} so that $\cW_2=\tfrac{\sqrt2}{144}\lamh^{-1/2}-\tfrac{\pi}{180}$ and $\cW_3=\tfrac{\sqrt2}{5184\pi^{2}}\lamh^{-3/2}-\tfrac{\pi}{3780}$. Each term of \eqref{eq:Wgen} descends from one term of the parent. The $\lamh$-dependent piece is the binomial expansion of the shifted $3/2$-power. The constant comes from the constant map: rescaling $y=kx$ in \eqref{eq:CW}, the large-$k$ expansion of $\log(2\cosh(y/k))$ and the moments $\int_0^\infty y^{2n}\log(1-\e^{-y})\,\rd y=-(2n)!\,\zeta(2n+2)$ generate the tower in $1/k^{2n}$, with one Bernoulli factor from each ingredient (Supplemental Material). The genus expansion thus inherits one geometric and one arithmetic ingredient from the two terms of the parent, every coefficient in closed form.

\begin{figure*}[htp!]
\centering

\begin{minipage}[t]{\columnwidth}
\vspace{0pt}
\centering
\includegraphics[width=\columnwidth]{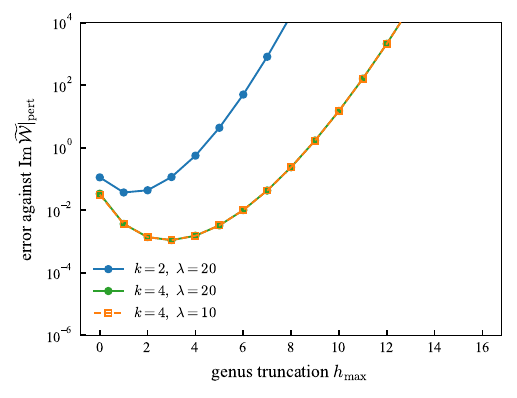}
\caption{The genus expansion is asymptotic. Plotted is the error left by
\eqref{eq:Wgenus} truncated at genus $h_{\max}$, for three choices of
$k$ and $\lambda$. The reference is the $800$-digit data with the
instanton series \eqref{eq:Wnp} subtracted. The error decreases to an
optimal order and then grows, as required by the factorial growth of
$\omega_n$.}
\label{fig:genus}
\end{minipage}
\hfill
\begin{minipage}[t]{\columnwidth}
\vspace{0pt}
\centering
\includegraphics[width=\columnwidth]{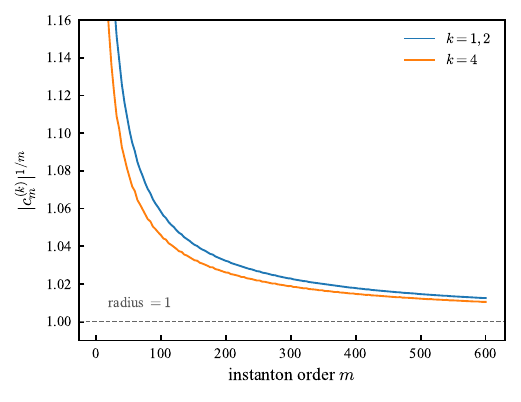}
\caption{The instanton series converges. The root test applied to
\eqref{eq:cm} shows both coefficient sequences approaching unity from
above. Polynomial bounds and the odd-prime subsequence give
$\limsup_{m\to\infty}|c_m^{(k)}|^{1/m}=1$, establishing unit radius of
convergence and absolute convergence of the physical remainder for
$0<q<1$.}
\label{fig:root}
\end{minipage}

\par\vspace{\floatsep}

\begin{minipage}[t]{\columnwidth}
\vspace{0pt}
\centering
\includegraphics[width=\columnwidth]{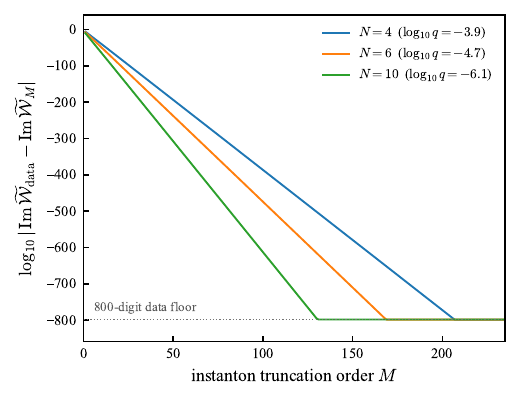}
\caption{Convergence and saturation at $k=4$. The residual between the
$800$-digit data and \eqref{eq:full}, truncated at $M$ instanton
sectors, falls geometrically over more than $770$ decades before
saturating the data floor near $10^{-798}$.}
\label{fig:conv}
\end{minipage}
\hfill
\begin{minipage}[t]{\columnwidth}
\vspace{0pt}
\centering
\includegraphics[width=\columnwidth]{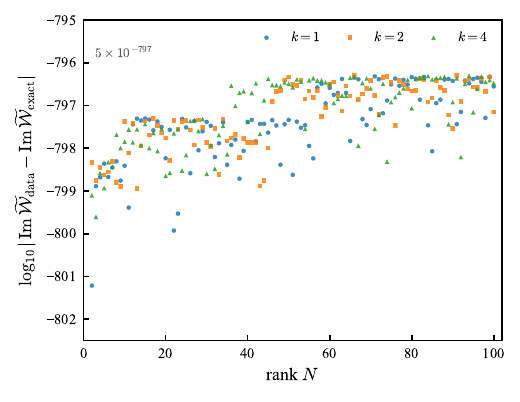}
\caption{Exactness at the special levels. For every $2\le N\le100$ at
$k=1,2,4$, the residual between the $800$-digit Bethe data and
\eqref{eq:full} lies between $10^{-801}$ and
$5.0\times10^{-797}$. All $297$ points therefore lie on the rounding
floor, with no outlier.}
\label{fig:exact}
\end{minipage}

\end{figure*}

The series they build is nevertheless asymptotic. Since $|B_{2n}|\sim2(2n)!/(2\pi)^{2n}$, the constants $\omega_n$ grow factorially, so \eqref{eq:Wgenus} is optimally truncated at a finite, $(k,\lambda)$-dependent genus [Fig.~\ref{fig:genus}, Supplemental Material], and is blind to effects nonperturbative in the type-IIA string coupling.

\emph{Exact modular completion.}---%
The exponentially small remainder is where $\Wt$ simplifies unexpectedly. At $k=1,2,4$ the entire finite-$N$ tail is a single convergent series,
\begin{align}
 \im\Wt(N,k)&=\frac{\pi^{2}}{3}\sqrt{\frac{k}{2}}\,\Nh^{3/2}+\CW(k)+\cW_{\rm np}(N,k) \, ,
\label{eq:full} \\
\cW_{\rm np}(N,k)&=\frac{k^{2}}{2\pi}\sum_{m\ge1}c^{(k)}_m\Big(t+\frac1m\Big)q^{m} \, ,
\label{eq:Wnp}
\end{align}
with one universal prefactor $k^{2}/2\pi$ and level-dependent coefficients
\be
c^{(k)}_m=\frac{(-1)^{m+1}}{m^{2}}\times
\begin{cases}
5\big[\sigma_3(m)-4\sigma_3(m/2)\big], & k=1,2 \, ,\\[2pt]
\sigma_3(m)-16\sigma_3(m/4), & k=4 \, .
\end{cases}
\label{eq:cm}
\ee

The divisor combinations in \eqref{eq:cm} are Fourier data of weight-four Eisenstein series. With $E_4(Q)=1+240\sum_{m\ge1}\sigma_3(m)Q^{m}$, and for any positive integer $L$,
\bea
E_4(Q)-L^{2}E_4(Q^{L})&=1-L^{2} \\
&\;+240\!\sum_{m\ge1}\Big[\sigma_3(m)-L^{2}\sigma_3\!\Big(\frac mL\Big)\Big]Q^{m} \, ,
\label{eq:E4}
\eea
so that $L=2$ delivers the coefficient sequence of \eqref{eq:cm} at $k=1,2$ and $L=4$ the one at $k=4$. Both are specific elements of the weight-four Eisenstein spaces on $\Gamma_0(2)$ and $\Gamma_0(4)$, of dimension $2$ and $3$~\cite{diamond2005first}, singled out by the data. The Chern--Simons levels and the modular levels track each other: $k=1,2\mapsto\Gamma_0(2)$ and $k=4\mapsto\Gamma_0(4)$. The decomposition $m^{-2}(t+1/m)=t/m^2+1/m^3$ identifies a first-order operator acting on an Eichler integral~\cite{pacsol2013modular}: weight four supplies the $m^{-3}$ primitive, while $tD$ produces the $t/m^2$ term. The relevant modular background is collected in the Supplemental Material. Write $D=Q\partial_Q$; after setting $Q=-q$, this is also $D=q\partial_q$. Define the Lambert series $\cL_3(Q)=\sum_{n\ge1}\Li_3(Q^{n})=\sum_{m\ge1}\sigma_3(m)m^{-3}Q^{m}$ for $|Q|<1$. It obeys $D^{3}\cL_3=(E_4-1)/240$ and is the coefficientwise Eichler primitive of the nonconstant part of $E_4$. With $D^{-3}$ understood to act only on nonconstant Fourier modes, the corresponding weight-four Eichler primitive of the combination in \eqref{eq:E4} is
\bea
\wt E_L(Q)&\equiv D^{-3}\big[E_4(Q)-L^{2}E_4(Q^{L})\big] \\
&=240\Big[\cL_3(Q)-\frac1L\cL_3(Q^{L})\Big] \, .
\label{eq:Etilde}
\eea
In terms of \eqref{eq:Etilde} the exact statement is then
\bea
\cW_{\rm np}(N,k)&=(1+tD)\,\Phi_k(q) \, \\
\Phi_{1}&=-\frac{1}{96\pi}\,\wt E_2(-q),\qquad \Phi_{2}=4\Phi_{1} \, , \\
\Phi_{4}&=-\frac{1}{30\pi}\,\wt E_4(-q) \, ,
\label{eq:Eichler}
\eea
since $(1+tD)$ turns a coefficient $a_m/m^{3}$ into $a_m(t/m^{2}+1/m^{3})$. Equation~\eqref{eq:Eichler} is an identity of formal $q$-series, equivalent to \eqref{eq:Wnp} term by term. The nome of the modular form is $-q$, not $q$---and that, not any choice of ours, is the origin of the alternating sign in \eqref{eq:cm}.

Two features deserve emphasis. First, $\sigma_3(m)/m^{2}=O(m^{1+\epsilon})$, while for every odd prime $\ell$ the subtraction term in \eqref{eq:cm} vanishes and $|c^{(k)}_{\ell}|\asymp\ell$. Hence $\limsup_{m\to\infty}|c^{(k)}_m|^{1/m}=1$. The two power series multiplying $t$ and the $m^{-1}$ term in \eqref{eq:Wnp} therefore have unit radius of convergence. In the physical domain $0<q<1$, where $t=-\log q$, the remainder is absolutely convergent [Fig.~\ref{fig:root}, Supplemental Material].

This is the sharpest contrast with $F_{\sthree}$, whose exponentially small sector is an asymptotic transseries of worldsheet, membrane and bound-state instantons~\cite{Drukker:2011zy,Hatsuda:2012dt,Hatsuda:2013gj}. Second, the scale is the one the string expects: with $\Nh/k=\lamh+2/(3k^{2})$, the exponent $t$ of \eqref{eq:shift} is the ABJM worldsheet-instanton action $2\pi\sqrt{2\lamh}$ of Ref.~\cite{Drukker:2010nc}, up to the $2/(3k^{2})$ carried by the shift. As for the levels, $k=1,2$ are the ones with enhanced $\cN=8$ supersymmetry~\cite{Gustavsson:2009pm}, which makes their common coefficient sequence natural; for $k=4$ we have no symmetry argument, and the shift of the modular level from $2$ to $4$ is an empirical fact that deserves an explanation.

The status of each statement is worth fixing here: \eqref{eq:E4} and \eqref{eq:Eichler} are analytic identities once \eqref{eq:cm} is given, whereas \eqref{eq:cm} itself is arithmetically reconstructed---exact rationals, identical when extracted independently from the $200$- and $800$-digit data---and established against the data rather than derived from the Bethe equations.

\emph{The discovery pipeline.}---%
The method is part of the result; Fig.~\ref{fig:pipe} of the Supplemental Material shows the workflow.

\emph{Data.} Precision is not a luxury but the whole game. At $k=1$, $N=800$ the $3/2$-power in \eqref{eq:parent} is of order $5\times10^{4}$, while the remainder it leaves behind is of order $10^{-107}$: reading the constant map means resolving structure some hundred and eleven orders of magnitude below the leading term, beyond the reach of double precision.

\emph{Physics-informed features.} The shift \eqref{eq:shift} is itself an output. Modelling each level as $a_k\,[N+s_k]^{3/2}+\CW(k)$ and solving for all three quantities from the largest ranks returns one shift per level; read against $s_k=-\alpha k-\beta/k$, these give $\alpha=1/24$ and $\beta=-2/3$ to some fifty digits, and the resulting $\Nh$ builds $t$ and $q$. The features for the constant map are the ansatz discussed above, tested rather than assumed. For the instanton sectors they are simply $q^m$ and $tq^m$: the divisor-sum structure of \eqref{eq:cm} is recognized only afterwards, from the reconstructed coefficients.

\emph{Exact recognition.} The two recognition problems need different machinery. A constant map is a \emph{single} number: there is nothing to regress, and the task is to decide whether it lies in the $\mathbb{Q}$-span of a given basis $\{b_1, b_2, \ldots, b_n\}$. Integer-relation detection (PSLQ)~\cite{ferguson1999analysis} answers exactly this, returning integers $a_0,a_1,\ldots,a_n$ with
$a_0\CW(k)+\sum_{j=1}^{n} a_j b_j=0$, or certifying that none exists below a height bound. A $q$-sector instead has many data points indexed by $N$ and few unknowns, so there we solve a high-precision linear system in the features above and reconstruct the coefficients as exact rationals. The output of that step for $k=1,2,4$ is the pair of integers $(5,-20)$ and $(1,-16)$ multiplying $\sigma_3(m)$ and $\sigma_3(m/2)$ or $\sigma_3(m/4)$---which is how \eqref{eq:cm}, and with it the Eisenstein structure, was found.

\emph{Rejection.} An overcomplete basis manufactures spurious relations, so the guards matter: the basis is first pruned of its own relations, and an accepted relation must have low height, survive precision reduction and basis deletion, and---decisively---predict data withheld from the fit: levels $k$ not used in assembling \eqref{eq:clausen-odd}--\eqref{eq:clausen-even} and ranks $N$ not used in fixing \eqref{eq:cm}. Discovery modules never read validation tables.

\emph{Quantitative validation.}---%
Equations~\eqref{eq:full}--\eqref{eq:cm} saturate the $800$-digit data. At fixed rank, the residual $|\im\Wt_{\rm data}-\im\Wt_{\rm exact}|$ falls geometrically with the instanton truncation order, its decay rate reproducing $\log_{10}q$ to better than $0.2\%$ over more than $770$ decades, before flattening onto the rounding floor of the data [Fig.~\ref{fig:conv}, Supplemental Material].
That floor is reached at every rank: across the whole range $2 \leq N \leq 100$ at $k=1,2,4$ the residual is at most $5.0\times10^{-797}$ [Fig.~\ref{fig:exact}, Supplemental Material]. The Clausen forms reproduce the plateaux at every $k\le50$ and agree numerically with both the Euler series and the integral \eqref{eq:CW} to at least sixty decimal places; the genus coefficients require no fit: \eqref{eq:W01} and \eqref{eq:Wgen} follow by expanding \eqref{eq:parent} at every $h$, and the truncated sums of \eqref{eq:Wgenus} are checked against the data in [Fig.~\ref{fig:genus}, Supplemental Material].

\emph{Comparison with the literature.}---%
The parent \eqref{eq:parent}
sharpens what is known: its scaling and shift are those established for the twisted index at the universal twist~\cite{Bobev:2022jte}. For $k=1,2,3,4$, its constant term agrees with the numerical estimates of Ref.~\cite{Bobev:2022wem}, obtained by fitting $\Wt(N,k)$ to a shifted $3/2$-power plus a constant. Their universal point $\Delta_a=1/2$ corresponds to $\Delta_a=\pi/2$ here, and the normalizations are related by $\CW(k)=2\pi\widehat g_0(k)$. The genus coefficients \eqref{eq:W01}--\eqref{eq:Wgen} are the twisted-superpotential counterparts of the $S^3$ genus expansion of Refs.~\cite{Marino:2009jd,Drukker:2010nc,Fuji:2011km}, with a different kernel and hence different $\cW_h$ and $\omega_n$. The finite Clausen formulas \eqref{eq:clausen-odd}--\eqref{eq:clausen-even}, the all-level integral representation \eqref{eq:CW}, the arbitrary-genus coefficients \eqref{eq:Wgen}, and the exact special-level completion \eqref{eq:full}--\eqref{eq:Eichler} are new.

The sharpest nonperturbative comparison is with the $S^3$ partition function. There, in the Fermi-gas formulation~\cite{Marino:2011eh}, an Airy-type contour transform of the grand potential $J(\mu,k)$ recovers the exact partition function, with $J$ built from the topological string on local $\mathbb{P}^1\times\mathbb{P}^1$: worldsheet instantons carry the unrefined free energy through its Gopakumar--Vafa invariants~\cite{Gopakumar:1998ii,Gopakumar:1998jq,Drukker:2010nc}, membrane instantons the refined string in the Nekrasov--Shatashvili limit~\cite{Marino:2011eh,Calvo:2012du,Hatsuda:2013oxa}, and their bound states enter through an effective chemical potential $\mu_{\mathrm{eff}}$~\cite{Hatsuda:2013gj}. Poles that individual sectors develop at special $k$ cancel between sectors, leaving the total finite at every physical level~\cite{Hatsuda:2012dt,Hatsuda:2013oxa,Hatsuda:2015gca}. The completion is thus intrinsically multi-sector.

At $k=1,2,4$, by contrast, the entire finite-$N$ remainder of $\Wt$---for the stated vacuum and branch---is a single absolutely convergent $q$-series: a first-order operator acting on an Eichler integral of a weight-four Eisenstein combination on $\Gamma_0(2)$ or $\Gamma_0(4)$. Quantum periods and spectral curves are familiar in the spectral-theory description of ABJM and local mirror symmetry~\cite{Kallen:2013qla,Grassi:2014zfa}; to our knowledge, a classical Eisenstein--Eichler structure surfacing directly in the on-shell twisted superpotential is new.

\emph{Discussion.}---%
The special-level result exposes an arithmetic organization of the ABJM twisted superpotential: the divisor combinations are Fourier coefficients of weight-four Eisenstein series on $\Gamma_0(2)$ and $\Gamma_0(4)$, and the $m^{-3}$ coefficients build the associated Eichler integrals. It is suggestive that the operator $(1+tD)$ involves $t=-\log q$, since an Eichler integral fails to be modular precisely by polynomial terms of this kind; but we have not derived how $(1+tD)\Phi_k$ transforms, and reading the universal twist as a projection onto a modular subsector remains an interpretation.

The dominant Cardy contribution to the superconformal index~\cite{Bhattacharya:2008zy} obeys $\log Z_{S^2\times S^1}\sim-\tfrac{\ii}{\omega}\im\Wt(N,k)$~\cite{Choi:2019dfu}. At $k=1,2,4$, Eqs.~\eqref{eq:full}--\eqref{eq:Wnp} therefore determine the complete coefficient of $\omega^{-1}$, including its nonperturbative finite-$N$ tail.

Holographically, at large $N$, the on-shell $\Wt$ controls the gravitational blocks that are glued into the entropy function of supersymmetric rotating, dyonic $\mathrm{AdS}_4$ black holes~\cite{Hosseini:2019iad}; the exponentials in \eqref{eq:Wnp} provide microscopic data that quantum corrections to this function would have to reproduce. Two questions thereby become concrete: whether Bethe vacua other than the dominant one contribute at the same exponential order, and which bulk configurations account for the terms in \eqref{eq:Wnp}. For the three-sphere, the analogous exponentials arise from Euclidean branes---fundamental strings wrapping a $\mathbb{CP}^1$~\cite{Drukker:2010nc} and D2-branes wrapping an $\mathbb{RP}^3$~\cite{Drukker:2011zy}---and whether the same objects account for \eqref{eq:Wnp} can now be tested against an explicit formula.

Several limitations bound our claims. Those of scope: \eqref{eq:Wnp} is established at $k=1,2,4$, and although the parent and the constant map hold at every integer level, we have no closed instanton series away from the exceptional ones, so whether a modular completion survives there is open; the data probe a single dominant Bethe vacuum in the branch and integer assignment fixed above, and we claim only that $\CW(k)$ is the quantity that saddle selects, not that it is invariant under all admissible branch changes; and \eqref{eq:Wgenus} is only the perturbative large-$k$, fixed-$\lambda$ expansion and does not include nonperturbative sectors. Those of method: both the constant map and the divisor coefficients \eqref{eq:cm} originate from high-precision reconstruction, validated as described above rather than derived from the Bethe equations; the general analytic reduction of the Euler sum to \eqref{eq:clausen-odd}--\eqref{eq:clausen-even} is deferred to Ref.~\cite{HosseiniTTI}. Deriving the finite-$N$ divisor structure---and, in particular, its Eisenstein organization---directly from the Bethe equations remains an open problem. Ref.~\cite{HosseiniTTI} will develop the twisted superpotential at greater length, alongside the parallel analysis of the topologically twisted index, where a constant map in closed form and a modular subsector appear as well.

\textit{Note added.}---%
While this work was being finalized, we became aware of the independent work~\cite{Hong:2026}. The Bethe-potential constant in Eq.~(29) therein is equivalent to our constant map in Eq.~\eqref{eq:CW}, with $C(k)=2\pi\widehat{g}_0(k,\Delta_{\mathrm{sc}})$. Closed-form expressions for the constant entering the twisted index were also obtained independently and will appear in our forthcoming companion paper~\cite{HosseiniTTI}.

\emph{Acknowledgments.}---%
It is a pleasure to thank Nikolay Bobev and Alberto Zaffaroni for discussions and for their careful reading of and comments on the draft.
Computations were performed using \texttt{Mathematica}, \texttt{mpmath}, and \texttt{SymPy}.
This research was supported by UK Research and Innovation (UKRI) under the UK government's Horizon Europe funding guarantee (Grant No. EP/Y027604/1).

\emph{Data availability.}---%
The data and code supporting the findings of this Letter are not publicly available and may be obtained from the author upon reasonable request.

\bibliography{twisted_superpotential_ML}

\begin{thebibliography}{42}%
\makeatletter
\providecommand \@ifxundefined [1]{%
 \@ifx{#1\undefined}
}%
\providecommand \@ifnum [1]{%
 \ifnum #1\expandafter \@firstoftwo
 \else \expandafter \@secondoftwo
 \fi
}%
\providecommand \@ifx [1]{%
 \ifx #1\expandafter \@firstoftwo
 \else \expandafter \@secondoftwo
 \fi
}%
\providecommand \natexlab [1]{#1}%
\providecommand \enquote  [1]{``#1''}%
\providecommand \bibnamefont  [1]{#1}%
\providecommand \bibfnamefont [1]{#1}%
\providecommand \citenamefont [1]{#1}%
\providecommand \href@noop [0]{\@secondoftwo}%
\providecommand \href [0]{\begingroup \@sanitize@url \@href}%
\providecommand \@href[1]{\@@startlink{#1}\@@href}%
\providecommand \@@href[1]{\endgroup#1\@@endlink}%
\providecommand \@sanitize@url [0]{\catcode `\\12\catcode `\$12\catcode
  `\&12\catcode `\#12\catcode `\^12\catcode `\_12\catcode `\%12\relax}%
\providecommand \@@startlink[1]{}%
\providecommand \@@endlink[0]{}%
\providecommand \url  [0]{\begingroup\@sanitize@url \@url }%
\providecommand \@url [1]{\endgroup\@href {#1}{\urlprefix }}%
\providecommand \urlprefix  [0]{URL }%
\providecommand \Eprint [0]{\href }%
\providecommand \doibase [0]{https://doi.org/}%
\providecommand \selectlanguage [0]{\@gobble}%
\providecommand \bibinfo  [0]{\@secondoftwo}%
\providecommand \bibfield  [0]{\@secondoftwo}%
\providecommand \translation [1]{[#1]}%
\providecommand \BibitemOpen [0]{}%
\providecommand \bibitemStop [0]{}%
\providecommand \bibitemNoStop [0]{.\EOS\space}%
\providecommand \EOS [0]{\spacefactor3000\relax}%
\providecommand \BibitemShut  [1]{\csname bibitem#1\endcsname}%
\let\auto@bib@innerbib\@empty
\bibitem [{\citenamefont {Pasquetti}(2012)}]{Pasquetti:2011fj}%
  \BibitemOpen
  \bibfield  {author} {\bibinfo {author} {\bibfnamefont {S.}~\bibnamefont
  {Pasquetti}},\ }\href {https://doi.org/10.1007/JHEP04(2012)120} {\bibfield
  {journal} {\bibinfo  {journal} {JHEP}\ }\textbf {\bibinfo {volume} {04}},\
  \bibinfo {pages} {120}},\ \Eprint {https://arxiv.org/abs/1111.6905}
  {arXiv:1111.6905 [hep-th]} \BibitemShut {NoStop}%
\bibitem [{\citenamefont {Beem}\ \emph {et~al.}(2014)\citenamefont {Beem},
  \citenamefont {Dimofte},\ and\ \citenamefont {Pasquetti}}]{Beem:2012mb}%
  \BibitemOpen
  \bibfield  {author} {\bibinfo {author} {\bibfnamefont {C.}~\bibnamefont
  {Beem}}, \bibinfo {author} {\bibfnamefont {T.}~\bibnamefont {Dimofte}},\ and\
  \bibinfo {author} {\bibfnamefont {S.}~\bibnamefont {Pasquetti}},\ }\href
  {https://doi.org/10.1007/JHEP12(2014)177} {\bibfield  {journal} {\bibinfo
  {journal} {JHEP}\ }\textbf {\bibinfo {volume} {12}},\ \bibinfo {pages}
  {177}},\ \Eprint {https://arxiv.org/abs/1211.1986} {arXiv:1211.1986 [hep-th]}
  \BibitemShut {NoStop}%
\bibitem [{\citenamefont {Nekrasov}\ and\ \citenamefont
  {Shatashvili}(2009)}]{Nekrasov:2009uh}%
  \BibitemOpen
  \bibfield  {author} {\bibinfo {author} {\bibfnamefont {N.~A.}\ \bibnamefont
  {Nekrasov}}\ and\ \bibinfo {author} {\bibfnamefont {S.~L.}\ \bibnamefont
  {Shatashvili}},\ }\href {https://doi.org/10.1016/j.nuclphysbps.2009.07.047}
  {\bibfield  {journal} {\bibinfo  {journal} {Nucl. Phys. B Proc. Suppl.}\
  }\textbf {\bibinfo {volume} {192-193}},\ \bibinfo {pages} {91} (\bibinfo
  {year} {2009})},\ \Eprint {https://arxiv.org/abs/0901.4744} {arXiv:0901.4744
  [hep-th]} \BibitemShut {NoStop}%
\bibitem [{\citenamefont {Benini}\ and\ \citenamefont
  {Zaffaroni}(2015)}]{Benini:2015noa}%
  \BibitemOpen
  \bibfield  {author} {\bibinfo {author} {\bibfnamefont {F.}~\bibnamefont
  {Benini}}\ and\ \bibinfo {author} {\bibfnamefont {A.}~\bibnamefont
  {Zaffaroni}},\ }\href {https://doi.org/10.1007/JHEP07(2015)127} {\bibfield
  {journal} {\bibinfo  {journal} {JHEP}\ }\textbf {\bibinfo {volume} {07}},\
  \bibinfo {pages} {127}},\ \Eprint {https://arxiv.org/abs/1504.03698}
  {arXiv:1504.03698 [hep-th]} \BibitemShut {NoStop}%
\bibitem [{\citenamefont {Choi}\ and\ \citenamefont
  {Hwang}(2020)}]{Choi:2019dfu}%
  \BibitemOpen
  \bibfield  {author} {\bibinfo {author} {\bibfnamefont {S.}~\bibnamefont
  {Choi}}\ and\ \bibinfo {author} {\bibfnamefont {C.}~\bibnamefont {Hwang}},\
  }\href {https://doi.org/10.1007/JHEP03(2020)068} {\bibfield  {journal}
  {\bibinfo  {journal} {JHEP}\ }\textbf {\bibinfo {volume} {03}},\ \bibinfo
  {pages} {068}},\ \Eprint {https://arxiv.org/abs/1911.01448} {arXiv:1911.01448
  [hep-th]} \BibitemShut {NoStop}%
\bibitem [{\citenamefont {Hosseini}\ and\ \citenamefont
  {Zaffaroni}(2022)}]{Hosseini:2022vho}%
  \BibitemOpen
  \bibfield  {author} {\bibinfo {author} {\bibfnamefont {S.~M.}\ \bibnamefont
  {Hosseini}}\ and\ \bibinfo {author} {\bibfnamefont {A.}~\bibnamefont
  {Zaffaroni}},\ }\href {https://doi.org/10.1007/JHEP12(2022)025} {\bibfield
  {journal} {\bibinfo  {journal} {JHEP}\ }\textbf {\bibinfo {volume} {12}},\
  \bibinfo {pages} {025}},\ \Eprint {https://arxiv.org/abs/2209.09274}
  {arXiv:2209.09274 [hep-th]} \BibitemShut {NoStop}%
\bibitem [{\citenamefont {Closset}\ \emph {et~al.}(2018)\citenamefont
  {Closset}, \citenamefont {Kim},\ and\ \citenamefont
  {Willett}}]{Closset:2018ghr}%
  \BibitemOpen
  \bibfield  {author} {\bibinfo {author} {\bibfnamefont {C.}~\bibnamefont
  {Closset}}, \bibinfo {author} {\bibfnamefont {H.}~\bibnamefont {Kim}},\ and\
  \bibinfo {author} {\bibfnamefont {B.}~\bibnamefont {Willett}},\ }\href
  {https://doi.org/10.1007/JHEP11(2018)004} {\bibfield  {journal} {\bibinfo
  {journal} {JHEP}\ }\textbf {\bibinfo {volume} {11}},\ \bibinfo {pages}
  {004}},\ \Eprint {https://arxiv.org/abs/1807.02328} {arXiv:1807.02328
  [hep-th]} \BibitemShut {NoStop}%
\bibitem [{\citenamefont {Hosseini}\ \emph {et~al.}(2019)\citenamefont
  {Hosseini}, \citenamefont {Hristov},\ and\ \citenamefont
  {Zaffaroni}}]{Hosseini:2019iad}%
  \BibitemOpen
  \bibfield  {author} {\bibinfo {author} {\bibfnamefont {S.~M.}\ \bibnamefont
  {Hosseini}}, \bibinfo {author} {\bibfnamefont {K.}~\bibnamefont {Hristov}},\
  and\ \bibinfo {author} {\bibfnamefont {A.}~\bibnamefont {Zaffaroni}},\ }\href
  {https://doi.org/10.1007/JHEP12(2019)168} {\bibfield  {journal} {\bibinfo
  {journal} {JHEP}\ }\textbf {\bibinfo {volume} {12}},\ \bibinfo {pages}
  {168}},\ \Eprint {https://arxiv.org/abs/1909.10550} {arXiv:1909.10550
  [hep-th]} \BibitemShut {NoStop}%
\bibitem [{\citenamefont {Aharony}\ \emph {et~al.}(2008)\citenamefont
  {Aharony}, \citenamefont {Bergman}, \citenamefont {Jafferis},\ and\
  \citenamefont {Maldacena}}]{Aharony:2008ug}%
  \BibitemOpen
  \bibfield  {author} {\bibinfo {author} {\bibfnamefont {O.}~\bibnamefont
  {Aharony}}, \bibinfo {author} {\bibfnamefont {O.}~\bibnamefont {Bergman}},
  \bibinfo {author} {\bibfnamefont {D.~L.}\ \bibnamefont {Jafferis}},\ and\
  \bibinfo {author} {\bibfnamefont {J.}~\bibnamefont {Maldacena}},\ }\href
  {https://doi.org/10.1088/1126-6708/2008/10/091} {\bibfield  {journal}
  {\bibinfo  {journal} {JHEP}\ }\textbf {\bibinfo {volume} {10}},\ \bibinfo
  {pages} {091}},\ \Eprint {https://arxiv.org/abs/0806.1218} {arXiv:0806.1218
  [hep-th]} \BibitemShut {NoStop}%
\bibitem [{\citenamefont {Marino}\ and\ \citenamefont
  {Putrov}(2010)}]{Marino:2009jd}%
  \BibitemOpen
  \bibfield  {author} {\bibinfo {author} {\bibfnamefont {M.}~\bibnamefont
  {Marino}}\ and\ \bibinfo {author} {\bibfnamefont {P.}~\bibnamefont
  {Putrov}},\ }\href {https://doi.org/10.1007/JHEP06(2010)011} {\bibfield
  {journal} {\bibinfo  {journal} {JHEP}\ }\textbf {\bibinfo {volume} {06}},\
  \bibinfo {pages} {011}},\ \Eprint {https://arxiv.org/abs/0912.3074}
  {arXiv:0912.3074 [hep-th]} \BibitemShut {NoStop}%
\bibitem [{\citenamefont {Drukker}\ \emph
  {et~al.}(2011{\natexlab{a}})\citenamefont {Drukker}, \citenamefont {Marino},\
  and\ \citenamefont {Putrov}}]{Drukker:2010nc}%
  \BibitemOpen
  \bibfield  {author} {\bibinfo {author} {\bibfnamefont {N.}~\bibnamefont
  {Drukker}}, \bibinfo {author} {\bibfnamefont {M.}~\bibnamefont {Marino}},\
  and\ \bibinfo {author} {\bibfnamefont {P.}~\bibnamefont {Putrov}},\ }\href
  {https://doi.org/10.1007/s00220-011-1253-6} {\bibfield  {journal} {\bibinfo
  {journal} {Commun. Math. Phys.}\ }\textbf {\bibinfo {volume} {306}},\
  \bibinfo {pages} {511} (\bibinfo {year} {2011}{\natexlab{a}})},\ \Eprint
  {https://arxiv.org/abs/1007.3837} {arXiv:1007.3837 [hep-th]} \BibitemShut
  {NoStop}%
\bibitem [{\citenamefont {Fuji}\ \emph {et~al.}(2011)\citenamefont {Fuji},
  \citenamefont {Hirano},\ and\ \citenamefont {Moriyama}}]{Fuji:2011km}%
  \BibitemOpen
  \bibfield  {author} {\bibinfo {author} {\bibfnamefont {H.}~\bibnamefont
  {Fuji}}, \bibinfo {author} {\bibfnamefont {S.}~\bibnamefont {Hirano}},\ and\
  \bibinfo {author} {\bibfnamefont {S.}~\bibnamefont {Moriyama}},\ }\href
  {https://doi.org/10.1007/JHEP08(2011)001} {\bibfield  {journal} {\bibinfo
  {journal} {JHEP}\ }\textbf {\bibinfo {volume} {08}},\ \bibinfo {pages}
  {001}},\ \Eprint {https://arxiv.org/abs/1106.4631} {arXiv:1106.4631 [hep-th]}
  \BibitemShut {NoStop}%
\bibitem [{\citenamefont {Marino}\ and\ \citenamefont
  {Putrov}(2012)}]{Marino:2011eh}%
  \BibitemOpen
  \bibfield  {author} {\bibinfo {author} {\bibfnamefont {M.}~\bibnamefont
  {Marino}}\ and\ \bibinfo {author} {\bibfnamefont {P.}~\bibnamefont
  {Putrov}},\ }\href {https://doi.org/10.1088/1742-5468/2012/03/P03001}
  {\bibfield  {journal} {\bibinfo  {journal} {J. Stat. Mech.}\ }\textbf
  {\bibinfo {volume} {1203}},\ \bibinfo {pages} {P03001} (\bibinfo {year}
  {2012})},\ \Eprint {https://arxiv.org/abs/1110.4066} {arXiv:1110.4066
  [hep-th]} \BibitemShut {NoStop}%
\bibitem [{\citenamefont {Bobev}\ \emph {et~al.}(2022)\citenamefont {Bobev},
  \citenamefont {Hong},\ and\ \citenamefont {Reys}}]{Bobev:2022jte}%
  \BibitemOpen
  \bibfield  {author} {\bibinfo {author} {\bibfnamefont {N.}~\bibnamefont
  {Bobev}}, \bibinfo {author} {\bibfnamefont {J.}~\bibnamefont {Hong}},\ and\
  \bibinfo {author} {\bibfnamefont {V.}~\bibnamefont {Reys}},\ }\href
  {https://doi.org/10.1103/PhysRevLett.129.041602} {\bibfield  {journal}
  {\bibinfo  {journal} {Phys. Rev. Lett.}\ }\textbf {\bibinfo {volume} {129}},\
  \bibinfo {pages} {041602} (\bibinfo {year} {2022})},\ \Eprint
  {https://arxiv.org/abs/2203.14981} {arXiv:2203.14981 [hep-th]} \BibitemShut
  {NoStop}%
\bibitem [{\citenamefont {Hatsuda}\ \emph
  {et~al.}(2013{\natexlab{a}})\citenamefont {Hatsuda}, \citenamefont
  {Moriyama},\ and\ \citenamefont {Okuyama}}]{Hatsuda:2012dt}%
  \BibitemOpen
  \bibfield  {author} {\bibinfo {author} {\bibfnamefont {Y.}~\bibnamefont
  {Hatsuda}}, \bibinfo {author} {\bibfnamefont {S.}~\bibnamefont {Moriyama}},\
  and\ \bibinfo {author} {\bibfnamefont {K.}~\bibnamefont {Okuyama}},\ }\href
  {https://doi.org/10.1007/JHEP01(2013)158} {\bibfield  {journal} {\bibinfo
  {journal} {JHEP}\ }\textbf {\bibinfo {volume} {01}},\ \bibinfo {pages}
  {158}},\ \Eprint {https://arxiv.org/abs/1211.1251} {arXiv:1211.1251 [hep-th]}
  \BibitemShut {NoStop}%
\bibitem [{\citenamefont {Calvo}\ and\ \citenamefont
  {Marino}(2013)}]{Calvo:2012du}%
  \BibitemOpen
  \bibfield  {author} {\bibinfo {author} {\bibfnamefont {F.}~\bibnamefont
  {Calvo}}\ and\ \bibinfo {author} {\bibfnamefont {M.}~\bibnamefont {Marino}},\
  }\href {https://doi.org/10.1007/JHEP05(2013)006} {\bibfield  {journal}
  {\bibinfo  {journal} {JHEP}\ }\textbf {\bibinfo {volume} {05}},\ \bibinfo
  {pages} {006}},\ \Eprint {https://arxiv.org/abs/1212.5118} {arXiv:1212.5118
  [hep-th]} \BibitemShut {NoStop}%
\bibitem [{\citenamefont {Hatsuda}\ \emph
  {et~al.}(2013{\natexlab{b}})\citenamefont {Hatsuda}, \citenamefont
  {Moriyama},\ and\ \citenamefont {Okuyama}}]{Hatsuda:2013gj}%
  \BibitemOpen
  \bibfield  {author} {\bibinfo {author} {\bibfnamefont {Y.}~\bibnamefont
  {Hatsuda}}, \bibinfo {author} {\bibfnamefont {S.}~\bibnamefont {Moriyama}},\
  and\ \bibinfo {author} {\bibfnamefont {K.}~\bibnamefont {Okuyama}},\ }\href
  {https://doi.org/10.1007/JHEP05(2013)054} {\bibfield  {journal} {\bibinfo
  {journal} {JHEP}\ }\textbf {\bibinfo {volume} {05}},\ \bibinfo {pages}
  {054}},\ \Eprint {https://arxiv.org/abs/1301.5184} {arXiv:1301.5184 [hep-th]}
  \BibitemShut {NoStop}%
\bibitem [{\citenamefont {Hatsuda}\ \emph {et~al.}(2014)\citenamefont
  {Hatsuda}, \citenamefont {Marino}, \citenamefont {Moriyama},\ and\
  \citenamefont {Okuyama}}]{Hatsuda:2013oxa}%
  \BibitemOpen
  \bibfield  {author} {\bibinfo {author} {\bibfnamefont {Y.}~\bibnamefont
  {Hatsuda}}, \bibinfo {author} {\bibfnamefont {M.}~\bibnamefont {Marino}},
  \bibinfo {author} {\bibfnamefont {S.}~\bibnamefont {Moriyama}},\ and\
  \bibinfo {author} {\bibfnamefont {K.}~\bibnamefont {Okuyama}},\ }\href
  {https://doi.org/10.1007/JHEP09(2014)168} {\bibfield  {journal} {\bibinfo
  {journal} {JHEP}\ }\textbf {\bibinfo {volume} {09}},\ \bibinfo {pages}
  {168}},\ \Eprint {https://arxiv.org/abs/1306.1734} {arXiv:1306.1734 [hep-th]}
  \BibitemShut {NoStop}%
\bibitem [{\citenamefont {Hatsuda}\ \emph {et~al.}(2015)\citenamefont
  {Hatsuda}, \citenamefont {Moriyama},\ and\ \citenamefont
  {Okuyama}}]{Hatsuda:2015gca}%
  \BibitemOpen
  \bibfield  {author} {\bibinfo {author} {\bibfnamefont {Y.}~\bibnamefont
  {Hatsuda}}, \bibinfo {author} {\bibfnamefont {S.}~\bibnamefont {Moriyama}},\
  and\ \bibinfo {author} {\bibfnamefont {K.}~\bibnamefont {Okuyama}},\ }\href
  {https://doi.org/10.1093/ptep/ptv145} {\bibfield  {journal} {\bibinfo
  {journal} {PTEP}\ }\textbf {\bibinfo {volume} {2015}},\ \bibinfo {pages}
  {11B104} (\bibinfo {year} {2015})},\ \Eprint
  {https://arxiv.org/abs/1507.01678} {arXiv:1507.01678 [hep-th]} \BibitemShut
  {NoStop}%
\bibitem [{\citenamefont {Udrescu}\ and\ \citenamefont
  {Tegmark}(2020)}]{Udrescu:2019mnk}%
  \BibitemOpen
  \bibfield  {author} {\bibinfo {author} {\bibfnamefont {S.-M.}\ \bibnamefont
  {Udrescu}}\ and\ \bibinfo {author} {\bibfnamefont {M.}~\bibnamefont
  {Tegmark}},\ }\href {https://doi.org/10.1126/sciadv.aay2631} {\bibfield
  {journal} {\bibinfo  {journal} {Sci. Adv.}\ }\textbf {\bibinfo {volume}
  {6}},\ \bibinfo {pages} {eaay2631} (\bibinfo {year} {2020})},\ \Eprint
  {https://arxiv.org/abs/1905.11481} {arXiv:1905.11481 [physics.comp-ph]}
  \BibitemShut {NoStop}%
\bibitem [{\citenamefont {Cranmer}(2023)}]{Cranmer:2023pysr}%
  \BibitemOpen
  \bibfield  {author} {\bibinfo {author} {\bibfnamefont {M.}~\bibnamefont
  {Cranmer}},\ }\href@noop {} {\bibfield  {journal} {\bibinfo  {journal} {arXiv
  e-prints}\ } (\bibinfo {year} {2023})},\ \Eprint
  {https://arxiv.org/abs/2305.01582} {arXiv:2305.01582 [astro-ph.IM]}
  \BibitemShut {NoStop}%
\bibitem [{\citenamefont {Ferguson}\ \emph {et~al.}(1999)\citenamefont
  {Ferguson}, \citenamefont {Bailey},\ and\ \citenamefont
  {Arno}}]{ferguson1999analysis}%
  \BibitemOpen
  \bibfield  {author} {\bibinfo {author} {\bibfnamefont {H.}~\bibnamefont
  {Ferguson}}, \bibinfo {author} {\bibfnamefont {D.}~\bibnamefont {Bailey}},\
  and\ \bibinfo {author} {\bibfnamefont {S.}~\bibnamefont {Arno}},\ }\href@noop
  {} {\bibfield  {journal} {\bibinfo  {journal} {Mathematics of Computation}\
  }\textbf {\bibinfo {volume} {68}},\ \bibinfo {pages} {351} (\bibinfo {year}
  {1999})}\BibitemShut {NoStop}%
\bibitem [{\citenamefont {Raayoni}\ \emph {et~al.}(2021)\citenamefont
  {Raayoni}, \citenamefont {Gottlieb}, \citenamefont {Manor}, \citenamefont
  {Pisha}, \citenamefont {Harris}, \citenamefont {Mendlovic}, \citenamefont
  {Haviv}, \citenamefont {Hadad},\ and\ \citenamefont
  {Kaminer}}]{Raayoni:2021ram}%
  \BibitemOpen
  \bibfield  {author} {\bibinfo {author} {\bibfnamefont {G.}~\bibnamefont
  {Raayoni}}, \bibinfo {author} {\bibfnamefont {S.}~\bibnamefont {Gottlieb}},
  \bibinfo {author} {\bibfnamefont {Y.}~\bibnamefont {Manor}}, \bibinfo
  {author} {\bibfnamefont {G.}~\bibnamefont {Pisha}}, \bibinfo {author}
  {\bibfnamefont {Y.}~\bibnamefont {Harris}}, \bibinfo {author} {\bibfnamefont
  {U.}~\bibnamefont {Mendlovic}}, \bibinfo {author} {\bibfnamefont
  {D.}~\bibnamefont {Haviv}}, \bibinfo {author} {\bibfnamefont
  {Y.}~\bibnamefont {Hadad}},\ and\ \bibinfo {author} {\bibfnamefont
  {I.}~\bibnamefont {Kaminer}},\ }\href
  {https://doi.org/10.1038/s41586-021-03229-4} {\bibfield  {journal} {\bibinfo
  {journal} {Nature}\ }\textbf {\bibinfo {volume} {590}},\ \bibinfo {pages}
  {67} (\bibinfo {year} {2021})}\BibitemShut {NoStop}%
\bibitem [{\citenamefont {Hosseini}()}]{HosseiniTTI}%
  \BibitemOpen
  \bibfield  {author} {\bibinfo {author} {\bibfnamefont {S.~M.}\ \bibnamefont
  {Hosseini}},\ }\bibinfo {note} {to appear}\BibitemShut {NoStop}%
\bibitem [{\citenamefont {Gustavsson}\ and\ \citenamefont
  {Rey}(2009)}]{Gustavsson:2009pm}%
  \BibitemOpen
  \bibfield  {author} {\bibinfo {author} {\bibfnamefont {A.}~\bibnamefont
  {Gustavsson}}\ and\ \bibinfo {author} {\bibfnamefont {S.-J.}\ \bibnamefont
  {Rey}},\ }\href@noop {} {\bibfield  {journal} {\bibinfo  {journal} {arXiv
  e-prints}\ } (\bibinfo {year} {2009})},\ \Eprint
  {https://arxiv.org/abs/0906.3568} {arXiv:0906.3568 [hep-th]} \BibitemShut
  {NoStop}%
\bibitem [{\citenamefont {Benini}\ \emph {et~al.}(2016)\citenamefont {Benini},
  \citenamefont {Hristov},\ and\ \citenamefont {Zaffaroni}}]{Benini:2015eyy}%
  \BibitemOpen
  \bibfield  {author} {\bibinfo {author} {\bibfnamefont {F.}~\bibnamefont
  {Benini}}, \bibinfo {author} {\bibfnamefont {K.}~\bibnamefont {Hristov}},\
  and\ \bibinfo {author} {\bibfnamefont {A.}~\bibnamefont {Zaffaroni}},\ }\href
  {https://doi.org/10.1007/JHEP05(2016)054} {\bibfield  {journal} {\bibinfo
  {journal} {JHEP}\ }\textbf {\bibinfo {volume} {05}},\ \bibinfo {pages}
  {054}},\ \Eprint {https://arxiv.org/abs/1511.04085} {arXiv:1511.04085
  [hep-th]} \BibitemShut {NoStop}%
\bibitem [{\citenamefont {Herzog}\ \emph {et~al.}(2011)\citenamefont {Herzog},
  \citenamefont {Klebanov}, \citenamefont {Pufu},\ and\ \citenamefont
  {Tesileanu}}]{Herzog:2010hf}%
  \BibitemOpen
  \bibfield  {author} {\bibinfo {author} {\bibfnamefont {C.~P.}\ \bibnamefont
  {Herzog}}, \bibinfo {author} {\bibfnamefont {I.~R.}\ \bibnamefont
  {Klebanov}}, \bibinfo {author} {\bibfnamefont {S.~S.}\ \bibnamefont {Pufu}},\
  and\ \bibinfo {author} {\bibfnamefont {T.}~\bibnamefont {Tesileanu}},\ }\href
  {https://doi.org/10.1103/PhysRevD.83.046001} {\bibfield  {journal} {\bibinfo
  {journal} {Phys. Rev. D}\ }\textbf {\bibinfo {volume} {83}},\ \bibinfo
  {pages} {046001} (\bibinfo {year} {2011})},\ \Eprint
  {https://arxiv.org/abs/1011.5487} {arXiv:1011.5487 [hep-th]} \BibitemShut
  {NoStop}%
\bibitem [{\citenamefont {Hosseini}(2026)}]{Hosseini:2025jxb}%
  \BibitemOpen
  \bibfield  {author} {\bibinfo {author} {\bibfnamefont {S.~M.}\ \bibnamefont
  {Hosseini}},\ }\href {https://doi.org/10.1103/2bpj-jf7x} {\bibfield
  {journal} {\bibinfo  {journal} {Phys. Rev. Lett.}\ }\textbf {\bibinfo
  {volume} {136}},\ \bibinfo {pages} {091601} (\bibinfo {year} {2026})},\
  \Eprint {https://arxiv.org/abs/2510.24837} {arXiv:2510.24837 [hep-th]}
  \BibitemShut {NoStop}%
\bibitem [{\citenamefont {Hosseini}\ and\ \citenamefont
  {Zaffaroni}(2016)}]{Hosseini:2016tor}%
  \BibitemOpen
  \bibfield  {author} {\bibinfo {author} {\bibfnamefont {S.~M.}\ \bibnamefont
  {Hosseini}}\ and\ \bibinfo {author} {\bibfnamefont {A.}~\bibnamefont
  {Zaffaroni}},\ }\href {https://doi.org/10.1007/JHEP08(2016)064} {\bibfield
  {journal} {\bibinfo  {journal} {JHEP}\ }\textbf {\bibinfo {volume} {08}},\
  \bibinfo {pages} {064}},\ \Eprint {https://arxiv.org/abs/1604.03122}
  {arXiv:1604.03122 [hep-th]} \BibitemShut {NoStop}%
\bibitem [{Note1()}]{Note1}%
  \BibitemOpen
  \bibinfo {note} {We normalize them by $x/(\protect \mathrm {e}^{x}-1)=\DOTSB
  \sum@ \slimits@ _{m\ge 0}B_m\protect \,x^{m}/m!$, so that $B_2=1/6$,
  $B_4=-1/30$ and $B_6=1/42$.}\BibitemShut {Stop}%
\bibitem [{\citenamefont {Diamond}\ and\ \citenamefont
  {Shurman}(2005)}]{diamond2005first}%
  \BibitemOpen
  \bibfield  {author} {\bibinfo {author} {\bibfnamefont {F.}~\bibnamefont
  {Diamond}}\ and\ \bibinfo {author} {\bibfnamefont {J.~M.}\ \bibnamefont
  {Shurman}},\ }\href@noop {} {\emph {\bibinfo {title} {A first course in
  modular forms}}},\ Vol.\ \bibinfo {volume} {228}\ (\bibinfo  {publisher}
  {Springer},\ \bibinfo {year} {2005})\BibitemShut {NoStop}%
\bibitem [{\citenamefont {Pa{\c{s}}ol}\ and\ \citenamefont
  {Popa}(2013)}]{pacsol2013modular}%
  \BibitemOpen
  \bibfield  {author} {\bibinfo {author} {\bibfnamefont {V.}~\bibnamefont
  {Pa{\c{s}}ol}}\ and\ \bibinfo {author} {\bibfnamefont {A.~A.}\ \bibnamefont
  {Popa}},\ }\href@noop {} {\bibfield  {journal} {\bibinfo  {journal}
  {Proceedings of the London Mathematical Society}\ }\textbf {\bibinfo {volume}
  {107}},\ \bibinfo {pages} {713} (\bibinfo {year} {2013})}\BibitemShut
  {NoStop}%
\bibitem [{\citenamefont {Drukker}\ \emph
  {et~al.}(2011{\natexlab{b}})\citenamefont {Drukker}, \citenamefont {Marino},\
  and\ \citenamefont {Putrov}}]{Drukker:2011zy}%
  \BibitemOpen
  \bibfield  {author} {\bibinfo {author} {\bibfnamefont {N.}~\bibnamefont
  {Drukker}}, \bibinfo {author} {\bibfnamefont {M.}~\bibnamefont {Marino}},\
  and\ \bibinfo {author} {\bibfnamefont {P.}~\bibnamefont {Putrov}},\ }\href
  {https://doi.org/10.1007/JHEP11(2011)141} {\bibfield  {journal} {\bibinfo
  {journal} {JHEP}\ }\textbf {\bibinfo {volume} {11}},\ \bibinfo {pages}
  {141}},\ \Eprint {https://arxiv.org/abs/1103.4844} {arXiv:1103.4844 [hep-th]}
  \BibitemShut {NoStop}%
\bibitem [{\citenamefont {Bobev}\ \emph {et~al.}(2023)\citenamefont {Bobev},
  \citenamefont {Choi}, \citenamefont {Hong},\ and\ \citenamefont
  {Reys}}]{Bobev:2022wem}%
  \BibitemOpen
  \bibfield  {author} {\bibinfo {author} {\bibfnamefont {N.}~\bibnamefont
  {Bobev}}, \bibinfo {author} {\bibfnamefont {S.}~\bibnamefont {Choi}},
  \bibinfo {author} {\bibfnamefont {J.}~\bibnamefont {Hong}},\ and\ \bibinfo
  {author} {\bibfnamefont {V.}~\bibnamefont {Reys}},\ }\href
  {https://doi.org/10.1007/JHEP02(2023)027} {\bibfield  {journal} {\bibinfo
  {journal} {JHEP}\ }\textbf {\bibinfo {volume} {02}},\ \bibinfo {pages}
  {027}},\ \Eprint {https://arxiv.org/abs/2210.15326} {arXiv:2210.15326
  [hep-th]} \BibitemShut {NoStop}%
\bibitem [{\citenamefont {Gopakumar}\ and\ \citenamefont
  {Vafa}(1998{\natexlab{a}})}]{Gopakumar:1998ii}%
  \BibitemOpen
  \bibfield  {author} {\bibinfo {author} {\bibfnamefont {R.}~\bibnamefont
  {Gopakumar}}\ and\ \bibinfo {author} {\bibfnamefont {C.}~\bibnamefont
  {Vafa}},\ }\href@noop {} {\bibfield  {journal} {\bibinfo  {journal} {arXiv
  e-prints}\ } (\bibinfo {year} {1998}{\natexlab{a}})},\ \Eprint
  {https://arxiv.org/abs/hep-th/9809187} {arXiv:hep-th/9809187} \BibitemShut
  {NoStop}%
\bibitem [{\citenamefont {Gopakumar}\ and\ \citenamefont
  {Vafa}(1998{\natexlab{b}})}]{Gopakumar:1998jq}%
  \BibitemOpen
  \bibfield  {author} {\bibinfo {author} {\bibfnamefont {R.}~\bibnamefont
  {Gopakumar}}\ and\ \bibinfo {author} {\bibfnamefont {C.}~\bibnamefont
  {Vafa}},\ }\href@noop {} {\bibfield  {journal} {\bibinfo  {journal} {arXiv
  e-prints}\ } (\bibinfo {year} {1998}{\natexlab{b}})},\ \Eprint
  {https://arxiv.org/abs/hep-th/9812127} {arXiv:hep-th/9812127} \BibitemShut
  {NoStop}%
\bibitem [{\citenamefont {Kallen}\ and\ \citenamefont
  {Marino}(2016)}]{Kallen:2013qla}%
  \BibitemOpen
  \bibfield  {author} {\bibinfo {author} {\bibfnamefont {J.}~\bibnamefont
  {Kallen}}\ and\ \bibinfo {author} {\bibfnamefont {M.}~\bibnamefont
  {Marino}},\ }\href {https://doi.org/10.1007/s00023-015-0421-1} {\bibfield
  {journal} {\bibinfo  {journal} {Annales Henri Poincare}\ }\textbf {\bibinfo
  {volume} {17}},\ \bibinfo {pages} {1037} (\bibinfo {year} {2016})},\ \Eprint
  {https://arxiv.org/abs/1308.6485} {arXiv:1308.6485 [hep-th]} \BibitemShut
  {NoStop}%
\bibitem [{\citenamefont {Grassi}\ \emph {et~al.}(2016)\citenamefont {Grassi},
  \citenamefont {Hatsuda},\ and\ \citenamefont {Marino}}]{Grassi:2014zfa}%
  \BibitemOpen
  \bibfield  {author} {\bibinfo {author} {\bibfnamefont {A.}~\bibnamefont
  {Grassi}}, \bibinfo {author} {\bibfnamefont {Y.}~\bibnamefont {Hatsuda}},\
  and\ \bibinfo {author} {\bibfnamefont {M.}~\bibnamefont {Marino}},\ }\href
  {https://doi.org/10.1007/s00023-016-0479-4} {\bibfield  {journal} {\bibinfo
  {journal} {Annales Henri Poincare}\ }\textbf {\bibinfo {volume} {17}},\
  \bibinfo {pages} {3177} (\bibinfo {year} {2016})},\ \Eprint
  {https://arxiv.org/abs/1410.3382} {arXiv:1410.3382 [hep-th]} \BibitemShut
  {NoStop}%
\bibitem [{\citenamefont {Bhattacharya}\ \emph {et~al.}(2008)\citenamefont
  {Bhattacharya}, \citenamefont {Bhattacharyya}, \citenamefont {Minwalla},\
  and\ \citenamefont {Raju}}]{Bhattacharya:2008zy}%
  \BibitemOpen
  \bibfield  {author} {\bibinfo {author} {\bibfnamefont {J.}~\bibnamefont
  {Bhattacharya}}, \bibinfo {author} {\bibfnamefont {S.}~\bibnamefont
  {Bhattacharyya}}, \bibinfo {author} {\bibfnamefont {S.}~\bibnamefont
  {Minwalla}},\ and\ \bibinfo {author} {\bibfnamefont {S.}~\bibnamefont
  {Raju}},\ }\href {https://doi.org/10.1088/1126-6708/2008/02/064} {\bibfield
  {journal} {\bibinfo  {journal} {JHEP}\ }\textbf {\bibinfo {volume} {02}},\
  \bibinfo {pages} {064}},\ \Eprint {https://arxiv.org/abs/0801.1435}
  {arXiv:0801.1435 [hep-th]} \BibitemShut {NoStop}%
\bibitem [{\citenamefont {Hong}()}]{Hong:2026}%
  \BibitemOpen
  \bibfield  {author} {\bibinfo {author} {\bibfnamefont {J.}~\bibnamefont
  {Hong}},\ }\bibinfo {note} {to appear}\BibitemShut {NoStop}%
\bibitem [{\citenamefont {Flajolet}\ and\ \citenamefont
  {Salvy}(1998)}]{FlajoletSalvy1998}%
  \BibitemOpen
  \bibfield  {author} {\bibinfo {author} {\bibfnamefont {P.}~\bibnamefont
  {Flajolet}}\ and\ \bibinfo {author} {\bibfnamefont {B.}~\bibnamefont
  {Salvy}},\ }\href {https://doi.org/10.1080/10586458.1998.10504356} {\bibfield
   {journal} {\bibinfo  {journal} {Experimental Mathematics}\ }\textbf
  {\bibinfo {volume} {7}},\ \bibinfo {pages} {15} (\bibinfo {year}
  {1998})}\BibitemShut {NoStop}%
\bibitem [{\citenamefont {Sofo}(2019)}]{Sofo2019}%
  \BibitemOpen
  \bibfield  {author} {\bibinfo {author} {\bibfnamefont {A.}~\bibnamefont
  {Sofo}},\ }\href {https://doi.org/10.1080/10652469.2019.1643851} {\bibfield
  {journal} {\bibinfo  {journal} {Integral Transforms and Special Functions}\
  }\textbf {\bibinfo {volume} {30}},\ \bibinfo {pages} {978} (\bibinfo {year}
  {2019})}\BibitemShut {NoStop}%
\end{thebibliography}%

\clearpage
\onecolumngrid
\begin{center}\large\textbf{Supplemental Material}\end{center}
\vspace{2mm}
\twocolumngrid

\emph{Constant-map integral: exact Euler reduction and numerical audit.}---%
For $k\in\mathbb{Z}_{>0}$ define
\bea
I_k&=\int_0^\infty\log(2\cosh x)\,\log\!\big(1-\e^{-kx}\big)\,\rd x \, ,\\
\Cint(k)&=-\frac{k^2\zeta(3)}{16\pi}+\frac{2k}{\pi}I_k \, .
\label{eq:sm-Ik}
\eea
Thus $\Cint(k)$ denotes the integral side of \eqref{eq:CW}, without presupposing its equality to the finite Clausen expression $\CW(k)$.
As $x\to0^+$, $\log(2\cosh x)=\log2+O(x^2)$ and
$\log(1-\e^{-kx})=\log(kx)+O(x)$, so the integrand is
$O(|\log x|)$. As $x\to\infty$, it is $O(x\e^{-kx})$.
Thus \eqref{eq:sm-Ik} is absolutely convergent for every positive integer
$k$.

\emph{Exact reduction to an Euler sum.}---%
With $z=\e^{-x}$,
\bea
\log(2\cosh x)&=-\log z+\log(1+z^2)\, ,\\
\log(1-\e^{-kx})&=\log(1-z^k) \, ,\qquad
\rd x=-\frac{\rd z}{z}\, .
\eea
Hence
\be
I_k=\int_0^1
\frac{\big[-\log z+\log(1+z^2)\big]\log(1-z^k)}{z}\,\rd z\, .
\label{eq:sm-zintegral}
\ee
For $0\le z<1$,
\bea
\log(1-z^k)&=-\sum_{n\ge1}\frac{z^{kn}}n,\\
\log(1+z^2)&=\sum_{m\ge1}\frac{(-1)^{m+1}z^{2m}}m.
\eea
The first expansion may be integrated term by term because
\be
\sum_{n\ge1}\frac1n\int_0^1z^{kn-1}|\log z|\,\rd z
=\frac{\zeta(3)}{k^2}<\infty \nn \, .
\ee
For the product of the two series,
\be
\sum_{m,n\ge1}\frac1{mn(2m+kn)}
\le\frac12\sum_{m,n\ge1}\frac1{m^{3/2}n^{3/2}}<\infty \nn \, ,
\ee
where $2m+kn\ge m+n\ge2\sqrt{mn}$. Fubini's theorem therefore
justifies the two interchanges and gives
\be
I_k=-\frac{\zeta(3)}{k^2}-S_k \, ,\qquad
S_k=\sum_{m,n\ge1}\frac{(-1)^{m+1}}{mn(2m+kn)} \, .
\label{eq:sm-double}
\ee
For $z\ge0$ define the generalized harmonic number by the
nonnegative integral
\be
\Hgen_z=\int_0^1\frac{1-x^z}{1-x}\,\rd x \, .
\label{eq:sm-Hdef}
\ee
Expanding $(1-x)^{-1}=\sum_{n\ge0}x^n$ and applying Tonelli's
theorem gives
\be
\Hgen_z=\sum_{n\ge1}\left(\frac1n-\frac1{n+z}\right)
=\psi(1+z)+\gamma \, ,
\label{eq:sm-Hseries}
\ee
where the last equality is the convergent series representation of the
digamma function $\psi$ and $\gamma$ is the
Euler--Mascheroni constant. The partial fraction
\be
\frac1{n(2m+kn)}=\frac1{2m}\left(\frac1n-\frac1{n+2m/k}\right) \nn
\ee
then yields
\be
S_k=\frac12\sum_{m\ge1}\frac{(-1)^{m+1}}{m^2}\,\Hgen_{2m/k} \, .
\label{eq:sm-euler}
\ee
If $N_m=\lceil2m/k\rceil$, then $2m/k\le N_m$. For $0<x<1$ the map $z\mapsto x^z$ is decreasing, and \eqref{eq:sm-Hdef} therefore gives $\Hgen_{2m/k}\le\Hgen_{N_m}=H_{N_m}$, where $H_N=\sum_{\ell=1}^N\ell^{-1}$. Since $H_N\le1+\log N$, we have $\Hgen_{2m/k}=O(\log m)$. The series in \eqref{eq:sm-euler} is consequently absolutely convergent. Combining the definition of $\Cint$, \eqref{eq:sm-double}, and \eqref{eq:sm-euler} gives the exact Euler representation
\be
\Cint(k)=-\frac{k^2\zeta(3)}{16\pi}
-\frac{2\zeta(3)}{\pi k}
-\frac{k}{\pi}\sum_{m\ge1}\frac{(-1)^{m+1}}{m^2}\,\Hgen_{2m/k} \, .
\label{eq:eulersum}
\ee

\emph{Status of the general evaluation.}---%
Everything through \eqref{eq:eulersum} is an exact consequence of the defining integral and holds for every positive integer $k$. To establish \eqref{eq:CW}, one must prove $\Cint(k)=\CW(k)$, namely that the integral side equals the parity-dependent Clausen expressions \eqref{eq:clausen-odd}--\eqref{eq:clausen-even}. Equivalently, one must evaluate
$\sum_{m\ge1}(-1)^{m+1}\Hgen_{2m/k}/m^2$
in a finite root-of-unity Clausen basis. The reduction requires a branch-controlled factorization at roots of unity, the evaluation of two Fourier kernels, and finite residue identities whose form depends on the parity of $k$. A complete proof, including the endpoint estimates, the exchanges of limits and integrals, and the finite residue reductions, is deferred to the companion paper~\cite{HosseiniTTI}; related rational-argument Euler sums are discussed in Refs.~\cite{FlajoletSalvy1998,Sofo2019}.

At $k=2$, \eqref{eq:eulersum} contains the ordinary harmonic numbers. Euler's classical identity
$\sum_{m\ge1}(-1)^{m+1}H_m/m^2=\frac58\zeta(3)$
gives $\Cint(2)=-5\zeta(3)/(2\pi)$, in agreement with \eqref{eq:clausen-even} at $p=1$, and proves $\Cint(2)=\CW(2)$ analytically. The other elementary entries of Table~\ref{tab:CW} follow by direct specialization of the finite Clausen formulas.

\emph{Independent numerical audit.}---%
We evaluated three quantities independently: $\CW_{\rm Cl}(k)$ from the finite Clausen expressions \eqref{eq:clausen-odd}--\eqref{eq:clausen-even}, $\CW_{\rm Euler}(k)$ from the right-hand side of \eqref{eq:eulersum}, and $\Cint(k)$ by direct quadrature of \eqref{eq:sm-Ik}. A \texttt{Mathematica} implementation based on \texttt{NSum} and \texttt{NIntegrate} was cross-checked by an independently written \texttt{mpmath} implementation. The latter was run at $70$ decimal digits and gives
\bea
\max_{1\le k\le50}
|\CW_{\rm Cl}(k)-\CW_{\rm Euler}(k)|
&<7\times10^{-69} \, , \\
\max_{1\le k\le50}
|\CW_{\rm Cl}(k)-\Cint(k)|
&<7\times10^{-69} \, .
\label{eq:sm-numerical-audit}
\eea
The largest displayed discrepancy occurs at $k=45$ and is at the working-precision floor. This comparison uses no Bethe-vacuum data and is logically separate from the plateau tests. It supplies strong numerical evidence for the general equivalence, but it does not replace the analytic proof deferred to Ref.~\cite{HosseiniTTI}.

\begin{table}[htp!]
\begin{ruledtabular}
\begin{tabular}{cl}
$k$ & $\CW(k)$\\
\colrule\smallstrut
$1$ & $-G-\frac{5\zeta(3)}{8\pi}$\\[2pt]
$2$ & $-\frac{5\zeta(3)}{2\pi}$\\[2pt]
$3$ & $G-\frac{19\zeta(3)}{8\pi}-\frac32\Cltwo\!\big(\frac{2\pi}{3}\big)$\\[2pt]
$4$ & $-\frac{3\zeta(3)}{\pi}$\\[2pt]
$5$ & $-G-\frac{253\zeta(3)}{40\pi}+\frac12\Cltwo\!\big(\frac{2\pi}{5}\big)+\frac72\Cltwo\!\big(\frac{4\pi}{5}\big)$\\[2pt]
$6$ & $-\frac{19\zeta(3)}{2\pi}+3\Cltwo\!\big(\frac{2\pi}{3}\big)$\\[2pt]
$7$ & $G-\frac{689\zeta(3)}{56\pi}-\frac52\Cltwo\!\big(\frac{2\pi}{7}\big)+\frac{11}{2}\Cltwo\!\big(\frac{4\pi}{7}\big)-\frac12\Cltwo\!\big(\frac{6\pi}{7}\big)$\\[2pt]
$8$ & $4G-\frac{31\zeta(3)}{2\pi}$\\
\end{tabular}
\end{ruledtabular}
\caption{The constant map $\CW(k)$ through $k=8$, evaluated from \eqref{eq:clausen-odd}--\eqref{eq:clausen-even}. Every entry is independently returned by the pipeline, as a certified integer relation on that level's own plateau. Catalan's constant enters at every odd $k$ through the explicit $(-1)^{(p+1)/2}G$ in \eqref{eq:clausen-odd}, and at even $k$ only when $8\mid k$, through $\Cltwo(\pi/2)$. The exceptional levels $k=1,2,4$ are elementary: at $k=1,2$ one has $p=1$ and the Clausen sums are empty, while at $k=4$ one has $p=2$ and the only Clausen value present is $\Cltwo(\pi)=0$.}
\label{tab:CW}
\end{table}

\emph{Genus coefficients from the parent.}---%
The Bernoulli tower in \eqref{eq:Wgen} descends from the constant map. After rescaling $y=kx$ in \eqref{eq:CW}, Watson's lemma applies to the exponentially decaying kernel $\log(1-\e^{-y})$. More explicitly, for any fixed truncation order one may split the $y$-integral at $y=\delta k$ with $0<\delta<\pi/2$: the tail is exponentially small in $k$, while Taylor's theorem on $0\le y\le\delta k$ bounds the omitted terms by the next even power of $y/k$. The resulting asymptotic expansion is generated by
\be
\log(2\cosh z)=\log2+\sum_{n\ge1}\frac{2^{2n}(2^{2n}-1)B_{2n}}{2n\,(2n)!}\,z^{2n}
\label{eq:coshexp}
\ee
and the moments $\int_0^\infty y^{2n}\log(1-\e^{-y})\,\rd y=-(2n)!\,\zeta(2n+2)$. The latter identity is obtained by expanding $-\log(1-\e^{-y})=\sum_{r\ge1}\e^{-ry}/r$, applying Tonelli's theorem to the nonnegative series, and evaluating $\int_0^\infty y^{2n}\e^{-ry}\,\rd y=(2n)!/r^{2n+1}$. Since $\zeta(2n+2)=\frac{(2\pi)^{2n+2}|B_{2n+2}|}{2(2n+2)!}$, the product $B_{2n}B_{2n+2}$ carries one factor from each expansion. The $\lamh$-dependent piece of \eqref{eq:Wgen} is the binomial expansion of the shifted $3/2$-power, as stated in the main text.

\emph{Modular background.}---%
A modular form of weight $w$ for a subgroup $\Gamma\subset\mathrm{SL}(2,\mathbb{Z})$ is a holomorphic function of $\tau$ in the upper half-plane satisfying $f\big(\tfrac{a\tau+b}{c\tau+d}\big)=(c\tau+d)^{w}f(\tau)$ for $\big(\begin{smallmatrix}a&b\\c&d\end{smallmatrix}\big)\in\Gamma$ and staying bounded at the \emph{cusps}---the boundary points $\mathbb{Q}\cup\{\ii\infty\}$, which $\Gamma$ groups into finitely many classes. Here $\Gamma=\Gamma_0(L)$, the matrices with $c$ divisible by $L$; it has two cusps for $L=2$ and three for $L=4$. Every such space is spanned by Eisenstein series, one per cusp when the weight is even and at least four, together with \emph{cusp forms}, which vanish at all cusps. At weight four these two levels admit no cusp forms at all~\cite{diamond2005first}, so the spaces are as small as they can be---dimension two on $\Gamma_0(2)$, three on $\Gamma_0(4)$---and \eqref{eq:E4} picks out one specific line in each. The Eichler integral is defined as follows. Let $f(\tau)=\sum_{m\ge0}a_mQ^m$, $Q=\e^{2\pi\ii\tau}$, be a modular form of even weight $w$, and $D=Q\partial_Q$, which multiplies the $m$-th Fourier coefficient by $m$. The coefficientwise Eichler primitive of $f$ is its $(w-1)$-fold formal antiderivative $D^{1-w}(f-a_0)=\sum_{m\ge1}a_m m^{1-w}Q^m$. For cusp forms its modular obstruction is the usual period polynomial of degree at most $w-2$; for an Eisenstein series with nonzero constant term, a transformation statement requires the corresponding regularized extension. We use only the unambiguous nonconstant-mode identity here and do not assume a transformation law for $(1+tD)\Phi_k$. For $w=4$ the coefficients are divided by $m^{3}$---the power seen in \eqref{eq:Wnp}. Finally, the Lambert-series identity used for $\cL_3$: the second equality follows by expanding $\Li_3(Q^n) = \sum_{r \geq 1} \frac{Q^{n r}}{r^3}$ and collecting $m = n r$: $\sum_{n \mid m} \frac{1}{(m/n)^3} = \frac{1}{m^3}\sum_{n \mid m} n^{3} = \frac{\sigma_3(m)}{m^3}$.

\emph{Supplementary figures.}---%
Figure~\ref{fig:pipe} summarizes the discovery workflow; Fig.~\ref{fig:genus} illustrates the asymptotic genus expansion \eqref{eq:Wgenus}; Fig.~\ref{fig:root} establishes the unit radius of convergence of \eqref{eq:Wnp}; Fig.~\ref{fig:conv} shows the geometric convergence of the truncated instanton series to the data floor at $k=4$; and Fig.~\ref{fig:exact} resolves the same saturation rank by rank across $k=1,2,4$, the pointwise form of the $5.0\times10^{-797}$ bound quoted in the main text.

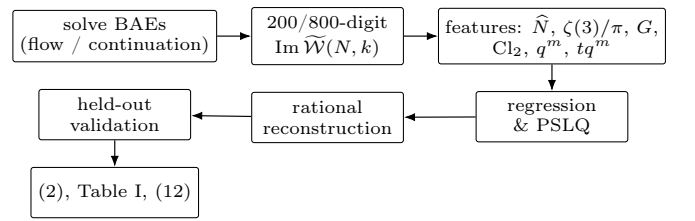
\begin{figure}[htp!]
\centering
\resizebox{\columnwidth}{!}{%
\begin{tikzpicture}[
  font=\scriptsize,
  node distance=3.6mm and 4.6mm,
  box/.style={
    draw,
    rounded corners=1pt,
    align=center,
    inner sep=2.6pt,
    minimum height=6.8mm,
    minimum width=20.5mm,
    line width=0.4pt
  },
  data/.style={box},
  phys/.style={box},
  ml/.style={box},
  aud/.style={box},
  ar/.style={-{Latex[length=1.5mm]},line width=0.45pt}
]
\node[data] (b) {solve BAEs\\(flow / continuation)};
\node[data,right=of b] (d) {$200$/$800$-digit\\$\im\Wt(N,k)$};
\node[phys,right=of d] (f) {features: $\Nh$, $\zeta(3)/\pi$, $G$,\\$\Cltwo$, $q^m$, $tq^m$};
\node[ml,below=of f] (r) {regression\\\& PSLQ};
\node[ml,below=of d] (p) {rational\\reconstruction};
\node[aud,below=of b] (v) {held-out\\validation};
\node[box,below=of v] (o)
  {\eqref{eq:shift}, Table~\ref{tab:CW}, \eqref{eq:cm}};

\draw[ar] (b) -- (d);
\draw[ar] (d) -- (f);
\draw[ar] (f) -- (r);
\draw[ar] (r) -- (p);
\draw[ar] (p) -- (v);
\draw[ar] (v) -- (o);
\end{tikzpicture}%
}
\caption{The machine-learning discovery pipeline. High-precision Bethe
vacua and physics-informed features feed integer-relation detection and
sparse regression; a candidate becomes a result only after held-out
validation. The pipeline returns the shift \eqref{eq:shift}, the certified
constant maps of Table~\ref{tab:CW}, and the divisor coefficients
\eqref{eq:cm}, which together assemble the closed form \eqref{eq:full}.}
\label{fig:pipe}
\end{figure}

\end{document}